\documentclass[11pt,a4paper]{article}
\usepackage{jheppub}

\newcommand{\bu}{{\bar u}}

\newcommand{\as}{\alpha_s}
\newcommand{\wh}{\widehat}
\newcommand{\nn}{\nonumber}

\newcommand{\ve}{\varepsilon}
\newcommand{\IM}{\mbox{\rm Im}}

\newcommand{\eqn}[1]{(\ref{#1})}

\newcommand{\mvs}{\vbox{\vskip 8mm}}

\newcommand{\MSb}{{\overline{\rm MS}}}

\newcommand{\sfrac}[2]{\mbox{$\frac{#1}{#2}$}}

\subheader{UAB-FT-702}

\title{The scalar gluonium correlator: large-\boldmath{$\beta_0$} and beyond}

\author{Matthias Jamin}

\affiliation{Instituci\'o Catalana de Recerca i Estudis Avan\c{c}ats (ICREA),\\
             IFAE, Theoretical Physics Group, UAB,
             E-08193 Bellaterra, Barcelona, Spain}

\emailAdd{jamin@ifae.es}

\abstract{The investigation of the scalar gluonium correlator is interesting
because it carries the quantum numbers of the vacuum and the relevant hadronic
current is related to the anomalous trace of the QCD energy-momentum tensor in
the chiral limit. After reviewing the purely perturbative corrections known up
to next-next-to-leading order, the behaviour of the correlator is studied to
all orders by means of the large-$\beta_0$ approximation. Similar to the QCD
Adler function, the large-order behaviour is governed by the leading
ultraviolet renormalon pole. The structure of infrared renormalon poles, being
related to the operator product expansion are also discussed, as well as a
low-energy theorem for the correlator that provides a relation to the
renormalisation group invariant gluon condensate, and the vacuum matrix
element of the trace of the QCD energy-momentum tensor.
}%

\keywords{QCD, gluonia, large-order behaviour, operator product expansion}

\begin{document}
\maketitle

\section{Introduction}\label{sect1}

The investigation of two-point correlation functions of hadronic currents plays
a very important role in phenomenological applications of QCD. The field gained
much impetus after the seminal work of Shifman, Vainshtein and Zakharov
\cite{svz79a,svz79b} in which mesonic correlation functions were studied.
Gluonic correlation functions were examined in refs.~\cite{nsvz80,nsvz81}, but
in subsequent years much less work went into the investigation of their
phenomenology, presumably because the situation on experimental data for the
spectra was and still is rather difficult. This is in particular so for the
scalar gluonium sector, which carries the quantum numbers of the vacuum and
where strong mixing with scalar meson states is expected. A selected set of
QCD sum rule analyses includes the works \cite{dp86,bs90,nar96,hs00,hkms08}
where also further references can be found. (See also the ``Note on scalar
mesons'' in the Review of Particle Physics \cite{pdg10}.)

From the theoretical perspective the investigation of the scalar gluonium
correlation function has several interesting aspects. First of all, the
respective current is given by the gluonic piece in the QCD Lagrangian. Next,
the corresponding renormalisation group invariant (RGI) current constitutes
the gluonic contribution to the anomalous trace of the QCD energy-momentum
tensor, one of the sources of the breaking of conformal symmetry \cite{cdj77}.
And finally, the expectation value of these currents with respect to the full
QCD vacuum represents the so-called {\em gluon~condensate}, one of the central
parameters in the framework of QCD sum rules. The gluon condensate then also
emerges in a low-energy theorem for the scalar gluonium correlator
\cite{nsvz81}.

The definition of the gluon condensate is plagued with an inherent ambiguity,
being related to {\em renormalon singularities} of the Borel transform of the
correlation function in the complex Borel plane \cite{mue85,mb98}. The
ambiguity arises because a prescription has to be chosen of how to treat the
{\em infrared} (IR) renormalon singularities on the positive real axis, in
order to define the Borel integral. On the other hand the IR renormalon
singularities are connected to the power corrections in the operator product
expansion (OPE) and their general structure can be analysed with the help of
the renormalisation group \cite{mb98,ben95,gn97,pin03,cgm03,bj08}. Still,
complete results for the Borel transform of two-point correlators are most
often only available in simplified models, like the large-$n_f$ or relatedly
the large-$\beta_0$ approximation, where $n_f$ is the number of fermion
flavours and $\beta_0$ the leading coefficient of the QCD $\beta$-function.
Nonetheless, valuable information on the structure of the correlators and their
large-order behaviour can be gleaned from such approximations. In the following
work, therefore, the above aspects shall be examined for the scalar gluonium
correlation function.
 
To begin, in section~2 the general structure of the perturbative series for
the scalar gluonium correlator will be discussed. To this end, two choices for
the hadronic current will be compared. The first consists of the gluonic term
in the QCD Lagrangian. However, as this current is not renormalisation group
invariant, the correlator cannot directly be connected to an observable
quantity. For this reason then a second, related correlator will be
investigated, which is constructed from the RGI combination that appears in the
trace of the QCD energy-momentum tensor. The spectral function (imaginary part)
of this correlator is an observable and also another physical quantity can be
constructed which bears similarities to the Adler function for the mesonic
vector correlator. The analytically available information on the perturbative
coefficients of these correlators will be reviewed.

In section~3, the large-$\beta_0$ approximation for the scalar gluonium
correlator will be computed and discussed. In particular the structure of the
ultraviolet and infrared renormalon poles will be investigated, which are
related to the large-order behaviour of the perturbative expansion of the
correlation function and the structure of higher-dimensional operator
corrections in the OPE. Furthermore, based on RG arguments, the structure of
the leading IR renormalon singularity will be derived beyond the large-$\beta_0$
approximation. Section~4 finally discusses a low-energy theorem for the scalar
gluonium correlator \cite{nsvz81} and its relation to the gluon condensate,
before the work is concluded in section~5.

\section{Scalar gluonium correlator}\label{sect2}

The most basic two-point correlation function that is relevant for the study
of scalar gluonium can be defined as
\begin{equation}
\label{TG2G2}
\Pi_{G^2}(q^2) \,\equiv\, i\!\int\!{\rm d}x\,{\rm e}^{iqx}\,
\langle\Omega|T\{J_G(x) J_G(0)\}|\Omega\rangle \,,
\end{equation}
where the gluonic current is given by
$J_G(x)\equiv G_{\mu\nu}^{\,a}(x)\,G^{\mu\nu\,a}(x)$, $G_{\mu\nu}^{\,a}(x)$
is the QCD field-strength tensor and $|\Omega\rangle$ represents the full
non-perturbative QCD vacuum.\footnote{Operator products at the same space-time
point are assumed to be renormalised by (modified) minimal subtraction in
dimensional regularisation.}

Analogous to the mesonic vector correlator (see e.g.~ref.~\cite{bj08}),
the purely perturbative part of $\Pi_{G^2}(q^2)$ has the following general
structure:
\begin{equation}
\label{PiG2}
\Pi_{G^2}^{\rm PT}(s) \,=\, -\,\frac{(N_c^2-1)}{4\pi^2}\,s^2
\sum\limits_{n=0}^\infty a_\mu^n \sum\limits_{k=0}^{n+1} c_{n,k}\,L^k  \,,
\quad L\,\equiv\, \ln\frac{-s}{\mu^2} \,,
\end{equation}
with $a_\mu\equiv a(\mu^2) \equiv\as(\mu)/\pi$ and $\mu$ the renormalisation
scale. $\Pi_{G^2}(s)$ itself is not a physical quantity. However, after
multiplying an appropriate function in the coupling that is related to the QCD
$\beta$-function and will be discussed in detail below, the spectral function
is physical, as well as the third derivative of $\Pi_{G^2}(s)$ with respect to
$s$, because for large $s$, $\Pi_{G^2}^{\rm PT}(s)$ scales as $s^2$ and three
subtractions are required. Let us first concentrate on the spectral function
$\rho_{G^2}(s)$, related to $\Pi_{G^2}(s)$,
\begin{equation}
\label{rhoG2}
\rho_{G^2}(s) \,\equiv\, \frac{1}{\pi}\,\IM\,\Pi_{G^2}(s+i0) \,.
\end{equation}

Up to second order in $\as$ the explicit expansion of the perturbative
spectral function $\rho_{G^2}^{\rm PT}(s)$ is found to be
\begin{equation}
\label{rhoG2ex}
\rho_{G^2}^{\rm PT}(s) \,=\, \frac{(N_c^2-1)}{4\pi^2}\,s^2 \Big[\,
c_{0,1} + (c_{1,1}+2c_{1,2}\bar L)\,a_\mu + (c_{2,1}-\pi^2 c_{2,3}+
2c_{2,2}\bar L+3c_{2,3}\bar L^2)\,a_\mu^2 \,\Big] \,,
\end{equation}
where $s\geq 0$ and $\bar L\equiv \ln(s/\mu^2)$. Like for the Adler function,
the perturbative coefficients $c_{n,0}$ are unphysical constants that have to
drop out for every physical quantity, the $c_{n,1}$ can be chosen as the only
independent coefficients, and, as will become evident from what follows, all
other $c_{n,k}$ for $k>1$ can be determined from the renormalisation group
equation (RGE). Comparing eq.~\eqn{rhoG2ex} with the analytic computation of
eq.~(10) of ref.~\cite{cks97} (for an earlier next-to-leading order calculation
see ref.~\cite{kkp82}), and setting $N_c=3$, the first three independent
coefficients are found to be:
\begin{equation}
\label{cn1}
c_{0,1} = 1 \,, \quad
c_{1,1} = \sfrac{73}{4} - \sfrac{7}{6}\,n_f \,, \quad
c_{2,1} = \sfrac{37631}{96} - \sfrac{495}{8}\,\zeta_3 - (\sfrac{7189}{144}-
\sfrac{5}{4}\,\zeta_3) n_f + \sfrac{127}{108}\,n_f^2 \,.
\end{equation}

In order to deduce constraints from the RGE, we have to work with a
renormalisation group invariant current. In the chiral limit, where the
operator $J_G(x)$ does not mix with $m_q\,\bar q(x)\,q(x)$ or $m_q^4$,
such a current can be chosen to be \cite{nt83}
\begin{equation}
\label{JGtilde}
\hat J_G(x) \,\equiv\, \frac{\beta(a)}{\beta_1 a}\,J_G(x) \,=\,
\frac{\beta(a)}{\beta_1 a}\,G_{\mu\nu}^{\,a}(x)\,G^{\mu\nu\,a}(x)
\,\approx\, a\,G_{\mu\nu}^{\,a}(x)\,G^{\mu\nu\,a}(x) \,.
\end{equation}
Here, the QCD $\beta$-function is defined as
\begin{equation}
\label{betafun}
\beta(a) \,\equiv\, -\,\mu\,\frac{{\rm d}a}{{\rm d}\mu} \,=\,
\beta_1\,a^2 + \beta_2\,a^3 + \ldots \,.
\end{equation}
Explicitely, at $N_c=3$, the first coefficient of the $\beta$-function reads
$\beta_1=11/2-n_f/3$. Let us remark that in the chiral limit $\hat J_G(x)$
is proportional to the trace of the QCD energy-momentum tensor
$\theta_\mu^{\,\mu}(x)$ \cite{nt83,sc88,dl91}, with the relation being given by
\begin{equation}
\label{thetamumu}
\theta_\mu^{\,\mu}(x) \,=\, -\,\frac{\beta(a)}{4\,a}\,G_{\mu\nu}^{\,a}(x)\,
G^{\mu\nu\,a}(x) \,=\, -\,\frac{\beta_1}{4}\,\hat J_G(x) \,.
\end{equation}

In analogy to $\Pi_{G^2}(q^2)$ of eq.~\eqn{TG2G2}, we can define the two-point
correlator for the current $\hat J_G(x)$, which expressed in terms of
$\Pi_{G^2}(q^2)$ takes the form:
\begin{equation}
\label{TG2G2tilde}
\wh\Pi_{G^2}(q^2) \,=\, \left(\frac{\beta(a)}{\beta_1 a}\right)^{\!2}
\Pi_{G^2}(q^2) \,.
\end{equation}
The spectral function $\hat\rho_{G^2}(s)\equiv\IM\,\wh\Pi_{G^2}(s+i0)/\pi$
corresponding to this correlator now is a RGI ``physical'' quantity, which
satisfies a homogeneous RGE. Working out the constraints from the RGE, up to
next-next-to-leading the following relations for the coefficients $c_{n,k}$
are found:
\begin{equation}
\label{cnk}
c_{1,2} \,=\, -\,\sfrac{\beta_1}{2}\,c_{0,1} \,, \qquad
c_{2,2} \,=\, -\,\beta_2\,c_{0,1} - \sfrac{3\beta_1}{4}\,c_{1,1} \,, \qquad
c_{2,3} \,=\,    \sfrac{\beta_1^2}{4}\,c_{0,1} \,.
\end{equation}
From the logarithmic terms of eq.~(10) of ref.~\cite{cks97} it can easily
be verified that these relations are satisfied.

The results given above can also be compared to corresponding expressions of
ref.~\cite{sho03}. Taking into account the fact that the definition of the
gluonic current in \cite{sho03} is $J_G^{\rm SHO}(x) = \pi\hat J_G(x)$, and
setting the number of quark flavours $n_f=3$, one obtains
\begin{eqnarray}
\label{a0to2}
a_0 &=& -\,2\,a^2 \left[\, 1 + \sfrac{659}{36}\,a + \left(
\sfrac{822569}{2592} - \sfrac{465}{8}\zeta_3 \right) a^2 \,\right] \,, \nn \\
\mvs
a_1 &=& \phantom{-}\,2\,a^3 \left[\, \sfrac{9}{4} +
\sfrac{2105}{32}\,a \,\right] \,, \qquad
a_2 \,=\, -\,\sfrac{81}{8}\,a^4 \,,
\end{eqnarray}
with $a_0$, $a_1$ and $a_2$ having been defined in eq.~(30) of
ref.~\cite{sho03}. The above expressions are found in agreement to eq.~(31)
of \cite{sho03}, there given partially numerically.

As has been noted before, since $\wh\Pi_{G^2}(s)$ behaves proportional to
$s^2$ for large $s$, it satisfies a dispersion relation with three
subtraction constants,
\begin{equation}
\label{PitildeDR}
\wh\Pi_{G^2}(s) \,=\, \wh\Pi_{G^2}(0) + s\,\wh\Pi_{G^2}^{'}(0) +
\frac{s^2}{2}\,\wh\Pi_{G^2}^{''}(0) +
s^3\!\int \frac{\hat\rho_{G^2}(s')}{(s')^3(s'-s-i0)}\,{\rm d}s' \,,
\end{equation}
where the prime denotes a derivative with respect to $s$. Therefore, besides
the spectral function $\hat\rho_{G^2}(s)$, taking three derivatives of
$\wh\Pi_{G^2}(s)$ another function $D_{G^2}(s)$ which is a physical quantity
can be defined in analogy to the QCD Adler function:
\begin{equation}
\label{DG2}
D_{G^2}(s) \,\equiv\, -\,s\,\frac{{\rm d}^3 \wh\Pi_{G^2}(s)}{{\rm d}s^3} \,.
\end{equation}
Employing eq.~\eqn{PiG2}, the general structure of the perturbative expansion 
for $D_{G^2}(s)$ is found to be
\begin{eqnarray}
\label{DG2PT}
D_{G^2}^{\rm PT}(s) &=& \frac{(N_c^2-1)}{4\pi^2}\,
\left(\frac{\beta(a_\mu)}{\beta_1 a_\mu}\right)^{\!2} \sum\limits_{n=0}^\infty
a_\mu^n \sum\limits_{k=1}^{n+1} k\,c_{n,k}\cdot \nn \\
\mvs
&& \cdot\Big[\, 2L^{k-1}+\,3(k-1)L^{k-2} + (k-1)(k-2)L^{k-3} \,\Big] \,.
\end{eqnarray}
Resumming the logarithms in $D_{G^2}^{\rm PT}(s)$ with the scale choice
$\mu^2=-s$, the general expansion as well as the perturbative expression up
to second order are found to be
\begin{eqnarray}
\label{DG2PTres}
D_{G^2}^{\rm PT}(s) &=& \frac{(N_c^2-1)}{2\pi^2}\,
\left(\frac{\beta(a(-s))}{\beta_1 a(-s)}\right)^{\!2} \sum\limits_{n=0}^\infty
\big[\, c_{n,1} + 3 c_{n,2} + 3 c_{n,3} \,\big]\,a(-s)^n  \nn \\
\mvs
&=& \frac{4}{\pi^2}\,a^2(-s) \biggl[\,
1 + \frac{161}{11}\,a(-s) + \biggl( \frac{1269361}{5808} - \frac{495}{8} \zeta_3
\biggr) a^2(-s) + \ldots \,\biggr] \,,
\end{eqnarray}
where in the second line the explicit expression for $n_f=0$ is provided. One
notices that the perturbative higher-order $\as$ corrections are large. To
obtain a reasonable behaviour of this perturbative series, one should have
$\alpha_s(-s)$ as small as roughly $0.12$ which requires scales as large as
about the $Z$-boson mass scale $M_Z$, and thus precludes a low-energy sum rule
analysis. Nonetheless, one may proceed to investigate the general properties
of the gluonium correlation function.

Since aspects of the perturbative series related to the renormalon ambiguity
of the gluon condensate shall be studied, it is also of interest to consider
the large-$\beta_0$ approximation (or large-$\beta_1$ approximation in my
notation, though I shall keep the historical terminology). In this
approximation the Adler-like function of eq.~\eqn{DG2} turns out to be:
\begin{equation}
\label{DG2lb0}
D_{G^2}^{{\rm large-}\beta_0}(s) \,=\, \frac{4}{\pi^2}\,a^2(-s) \left[\, 1 +
2\,\beta_1\,a(-s) + \sfrac{83}{24}\,\beta_1^2\,a^2(-s) + \ldots \,\right] \,.
\end{equation}
Numerically, the large-$\beta_0$ approximation is reasonably close to the full
result, $11$ versus $14.6$ at order $\as$ and $104.6$ versus $144.2$ at order
$\as^2$, so that it appears to make sense to work out the complete
large-$\beta_0$ approximation to $D_{G^2}^{\rm PT}(s)$. The behaviour of
the series in large-$\beta_0$ is slightly better than in full QCD as the
perturbative coefficients are somewhat smaller. The large-$\beta_0$
approximation to all perturbative orders shall be developed in the next
section.

\section{Large-{\boldmath{$\beta_0$}} approximation}\label{sect3}

Like in the case of mesonic correlation functions, the large-$\beta_0$
approximation can be calculated by first computing fermion-chain diagrams that
correspond to the limit of a large number of fermion flavours $n_f$ and then
applying the so-called {\em naive non-abelianisation}, that is replacing the
$n_f$ dependence by the leading coefficient of the QCD $\beta$-function,
which then also incorporates a gauge invariant set of gluon-loop diagrams
\cite{bg94,bb94}. In practice, one may either explicitly calculate the required
fermion-chain diagrams for $\Pi_{G^2}$ or proceed directly to a computation of
the Borel transform of $\Pi_{G^2}$ or $D_{G^2}$. As below the structure of the
Borel transform $B[D_{G^2}(s)](u)$ will be discussed in relation to the OPE,
first of all this Borel transform shall be presented. A discussion of the
large-$\beta_0$ result for $\Pi_{G^2}(s)$ is relegated to appendix~A.

Technically, the computation of the large-$\beta_0$ approximation for the
gluonium correlator proceeds through inserting chains of fermion loops in
both gluon propagators of the lowest order gluon loop diagram. Employing the
Dyson-resummed form of the gluon propagator, together with the convolution
theorem for the Borel transform (see e.g. eq.~(2.8) of ref.~\cite{bs96}),
one arrives at
\begin{eqnarray}
\label{BDG2u}
B[D_{G^2}(s)](u) &=& \frac{3(N_c^2-1)}{\pi^3\beta_1}\,\biggl(\frac{-s}{\mu^2}
\,{\rm e}^C\biggr)^{\!-u} \,\frac{\Gamma(1+u)}{\Gamma(4-u)} \cdot \nn \\
\mvs
&& \cdot \int\limits_0^u \!{\rm d}\bu\, \big[ 2-u+\bu(u-\bu) \big]\,
\frac{\Gamma(2-\bu)\Gamma(2-u+\bu)}{\Gamma(1+\bu)\Gamma(1+u-\bu)} \,,
\end{eqnarray}
where $C$ is a scheme-dependent constant which in the $\MSb$ scheme takes the
value $-5/3$. Assuming a positive coupling $\alpha_s$, the relation between the
Borel transform $B[D_{G^2}](u)$ and the Adler-like function $D_{G^2}$ is given
by
\begin{equation}
\label{Btrafo}
D_{G^2}(a) \,=\, \frac{2\pi}{\beta_1} \int\limits_0^\infty \!{\rm d}u\,
{\rm e}^{-\frac{2u}{\beta_1a}} B[D_{G^2}](u) \,.
\end{equation}
Without further specification, the Borel-integral on the right-hand side is
only well defined if $B[D_{G^2}](u)$ has no poles or cuts on the positive real
axis. Similarly to the case of the QCD Adler function, as will be discussed
in more detail further below, this is not the case for the expression of 
eq.~\eqn{BDG2u}. Still, the perturbative expansion, corresponding to an
expansion of $B[D_{G^2}](u)$ in powers of $u$, is well defined, and takes the
form
\begin{eqnarray}
\label{BDG2uexp}
B[D_{G^2}(s)](u) &\equiv& \frac{(N_c^2-1)}{\pi^3\beta_1}\;
\sum\limits_{n=0}^\infty \,a_n\,u^{n+1} \nn \\
\mvs
&=& \frac{(N_c^2-1)}{\pi^3\beta_1}\,\left[\,
u + 2\,u^2 + \sfrac{83}{36}\, u^3 + \big(\sfrac{1403}{648}-
\sfrac{\zeta_3}{3}\big) u^4 + \ldots\,\right] \,,
\end{eqnarray}
where in the second line the first few orders have been given explicitly.
Higher-order expansion coefficients $a_n$ are collected in table~\ref{tab1}
in numerical form up to the 21st order. Together with the formula
\begin{equation}
\int\limits_0^\infty \!{\rm d}u\,u^n\,{\rm e}^{-\frac{2u}{\beta_1 a}} \,=\,
n!\,\biggl(\frac{\beta_1 a}{2}\biggr)^{\!{n+1}} \,,
\end{equation}
leading to the perturbative expansion
\begin{equation}
\label{Dlargeb0}
D_{G^2}^{{\rm large-}\beta_0}(s) \,=\, \frac{(N_c^2-1)}{2\pi^2}\,a^2(-s)\,
\sum\limits_{n=0}^\infty \,\frac{(n+1)!}{2^n}\,a_n\,[\beta_1 a(-s)]^n \,,
\end{equation}
it is a straightforward matter to recover eq.~\eqn{DG2lb0} for $D_{G^2}(a)$
in the large-$\beta_0$ limit. Eq.~\eqn{Dlargeb0} also clearly displays the
factorially divergent asymptotic behaviour of the perturbative series.

\begin{table}[t]
\begin{center}
\begin{tabular}{rrrrrrr}
\hline\hline
$a_{0}\quad$ & $a_{1}\quad$ & $a_{2}\quad$ & $a_{3}\quad$ & $a_{4}\quad$ &
$a_{5}\quad$ & $a_{6}\quad$ \\
\hline
$1\quad$ & $2\quad$ & $2.30556$ & $1.76444$ & $1.14045$ & $0.55870$ &
$0.30433$ \\
\hline\hline
$a_{7}\quad$ & $a_{8}\quad$ & $a_{9}\quad$ & $a_{10}\quad$ & $a_{11}\quad$ &
$a_{12}\quad$ & $a_{13}\quad$ \\
\hline
$0.09359$ & $0.07845$ & $-\,0.00719$ & $0.03581$ & $-\,0.02458$ & $0.02891$ &
$-\,0.02726$ \\
\hline\hline
$a_{14}\quad$ & $a_{15}\quad$ & $a_{16}\quad$ & $a_{17}\quad$ & $a_{18}\quad$ &
$a_{19}\quad$ & $a_{20}\quad$ \\
\hline
$0.02789$ & $-\,0.02764$ & $0.02774$ & $-\,0.02770$ & $0.02772$ &
$-\,0.02771$ & $0.02771$ \\
\hline\hline
\end{tabular}
\caption{Numerical expansion coefficients $a_{n}$ of eq.~\eqn{BDG2uexp} up to
the 21st order.\label{tab1}}
\end{center}
\end{table}

Next, the singularity structure in the Borel plane, also known as renormalons,
shall be investigated. Via the factor $\Gamma(1+u)$, $B[D_{G^2}](u)$ has
ultraviolet (UV) renormalons for all negative integer $u$. The UV renormalon
singularity at $u=-1$ is also the closest to the origin and thus dominates the
perturbative series at large orders. In the vicinity of $u=-1$, the behaviour
of the Borel transform reads
\begin{equation}
B[D_{G^2}](u) \,\stackrel{u\to-1}{=}\, -\,\frac{(N_c^2-1)}{\pi^3\beta_1}
\,\frac{\sfrac{493}{3360}\,{\rm e}^C}{(1+u)} \,,
\end{equation}
from which the perturbative coefficients at large orders can be deduced. In
the $\MSb$ scheme, dominance of the leading UV renormalon singularity sets
in at about the 12th order, and inspecting table~\ref{tab1}, it is found to be
in accord with the behaviour of the coefficients $a_n$ at sufficiently large
$n$.

The singularities at positive $u$, the infrared (IR) renormalons, are connected
to the power corrections in the OPE. For physical correlation functions they
are generally expected to start at $u=2$ which corresponds to the dimension-4
gluon condensate. In the case at hand, however, the correlator $\Pi_{G^2}(s)$
itself is of dimension four, and therefore the gluon condensate only appears
as a subtraction constant which vanishes for the physical correlator $D_{G^2}$.
This is also seen from the fact that the Borel transform $B[D_{G^2}](u)$ stays
finite at $u=2$, and will be further discussed in connection to a low-energy
theorem for $\Pi_{G^2}$ in the next section. Hence, potential IR renormalon
singularities only start at integer $u\geq 3$.

Even though from RG arguments generally power-like IR renormalon singularities
are expected \cite{mb98,ben95,gn97,pin03,cgm03,bj08}, in large-$\beta_0$ for
$B[D_{G^2}](u)$ they turn out to be only logarithmic. To understand the origin
of this behaviour, first the generic case shall be investigated. The derivation
is completely analogous to the one for example presented in section~5 of
ref.~\cite{bj08}, the only difference being that now the correlator has an
additional multiplicative factor of $\as^\delta$, with $\delta=2$ for $D_{G^2}$.
The structure of an IR renormalon pole corresponding to an operator $O_d$ of
dimension $d$ is then found to be
\begin{equation}
\label{IRpole}
B[D_{G^2}^{\rm IR}](u) \,=\, \frac{d_p^{\rm IR}}{(p-u)^{1+\tilde\gamma}}\,
\big[\, 1 + {\cal O}(p-u) \,\big] \,,
\end{equation}
where $d_p^{\rm IR}$ are normalisations of the renormalon poles that cannot be
determined from the renormalisation group, and
\begin{equation}
p \,=\, \frac{d}{2} \,, \quad
\tilde\gamma \,=\, 2p\,\frac{\beta_2}{\beta_1^2} -
\frac{\gamma_{O_d}^{(1)}}{\beta_1} - \delta \,.
\end{equation}
Here, $\gamma_{O_d}^{(1)}$ is the leading order anomalous dimension of the
operator $O_d$.

At $u=3$, the only contributing operator corresponds to the dimension-6
gluonic condensate $\langle\Omega|gf^{abc}G_{\mu\nu}^a G_{\;\lambda}^{\,\nu,b}
G^{\lambda\mu,c}|\Omega\rangle$, whose leading anomalous dimension was
calculated to be $\gamma_{gG^3}^{(1)}=(7N_c+2n_f)/6$ in \cite{nt83,mor84}.
Hence, in the absence of quarks, that is $n_f=0$, the exponent in the
denominator of eq.~\eqn{IRpole} turns out to be $1+\tilde\gamma=108/121$
(independent of $N_c$). In the large-$\beta_0$ limit, on the other hand,
the $\beta_2$-term vanishes and $\gamma_{gG^3}^{(1)}/\beta_1=-1$. Thus, the
exponent $1+\tilde\gamma$ is found identically zero. This is reflected in the
fact that close to $u=3$ the Borel transform \eqn{BDG2u} behaves as
\begin{equation}
B[D_{G^2}](u) \,\stackrel{u\to 3}{=}\, -\,\frac{(N_c^2-1)}{\pi^3\beta_1}
\,3\,{\rm e}^{-3C}\,\ln\left[1-\sfrac{u}{3}\right] \,,
\end{equation}
only displaying a logarithmic singularity. Such a logarithmic behaviour of
the leading IR renormalon was also observed in ref.~\cite{cgm03}. A further
consequence of the softening of the leading IR renormalon singularity is the
fact that at no order it dominates the perturbative coefficients. This is in
contrast to the QCD Adler function for which in large-$\beta_0$ the leading
IR singularity at $u=2$ gives the most important contribution at intermediate
orders before the large-order behaviour governed by the UV singularity at
$u=-1$ sets in \cite{mb98}.\footnote{The latter behaviour is also expected
for the Adler function in full QCD \cite{bj08}.}

At positive integer $p\geq 4$, the Borel transform $B[D_{G^2}](u)$ again
behaves logarithmically like $\ln[1-u/p]$, but this time, due to the factor
$1/\Gamma(4-u)$ in eq.~\eqn{BDG2u}, these logarithms are further suppressed
by an additional factor $(p-u)$. Thus, at positive integer $u\geq 4$, the
Borel transform does not posses further singularities, but just the appearance
of new cuts which reflect the presence of higher-dimensional contributions in
the OPE. In order to properly define the Borel integral of eq.~\eqn{Btrafo},
hence a prescription should be chosen which circumvents the singularity at
$u=3$ and the cuts at all integer $u\geq 3$. This prescription then also gives
a meaning to the QCD condensate parameters.

\section{Low-energy theorem}\label{sect4}

Even though $\wh\Pi_{G^2}(s)$, and therefore also $\wh\Pi_{G^2}(0)$, are not
physical quantities, the question arises if a closed expression can be given
for the subtraction constant $\wh\Pi_{G^2}(0)$ of the dispersion relation
\eqn{PitildeDR}, similar to the case of the pseudoscalar correlation function
$\Psi_5(s)$ to be discussed further below. Such relations are termed low-energy
theorems (LET) and they are interesting as they provide additional constraints
on the theory in question. In fact, already in the seminal work \cite{nsvz81},
a LET for $\wh\Pi_{G^2}(s)$ was derived which reads:
\begin{equation}
\label{PiG2LET}
\wh\Pi_{G^2}(0) \,=\, \lim_{s\to 0}\, \wh\Pi_{G^2}(s) \,=\,
\frac{16}{\beta_1}\,\langle\Omega|\hat J_G|\Omega\rangle \,.
\end{equation}
The LET can also be expressed in terms of the correlator $\Pi_{\theta}(s)$
for the trace of the energy-momentum tensor $\theta_\mu^{\,\mu}(x)$ of
eq.~\eqn{thetamumu}, in which case it assumes the particularly simple form
\begin{equation}
\label{PithetaLET}
\Pi_{\theta}(0) \,=\, \lim_{s\to 0}\, \Pi_{\theta}(s) \,=\,
-\,4\,\langle\Omega|\theta_\mu^{\,\mu}|\Omega\rangle \,.
\end{equation}
The gluon condensate which appears on the rhs of \eqn{PiG2LET} and
\eqn{PithetaLET} has a renormalon ambiguity. This is for example seen in the
large-$\beta_0$ approximation for the Borel transform of $\wh\Pi_{G^2}(s)$
which, in contrast to $B[D_{G^2}](u)$, has a UV renormalon pole at $u=2$.
(See also appendix~A.) This pole provides an ambiguity of order
$\Lambda_{\rm QCD}^4$ which cancels against a corresponding ambiguity in the
definition of the gluon condensate. Nevertheless, this does not imply that
the gluon condensate is an observable, since the subtraction constant is not
directly related to the spectrum of $\wh\Pi_{G^2}(s)$.

The situation is analogous to the case of the pseudoscalar correlator
$\Psi_5(s)$. Here the subtraction constant $\Psi_5(0)$ is related to the quark
condensate $\langle\Omega|\bar qq|\Omega\rangle$, and the low-energy theorem
can also be derived for a hadronic representation of the currents which yields
the Gell-Mann-Oakes-Renner (GMOR) relation \cite{gmor68}. The derivation of the
GMOR relation can be generalised in the framework of chiral perturbation theory
($\chi$PT) \cite{gl85}, and the fact that the subtraction constant $\Psi_5(0)$
is not a physical observable then manifests itself through the appearance of
the unphysical $\chi$PT coupling $H_2$ at higher orders in the $\chi$PT
expansion \cite{mj02}, reflecting the dependence on the short-distance
renormalisation.

It might be perhaps interesting to study a GMOR-like relation also for the
scalar gluonium correlator $\wh\Pi_{G^2}(s)$. The framework for such an
analysis is available from the work of Donoghue and Leutwyler \cite{dl91}, who
derived chiral expansions for correlation functions of the energy-momentum
tensor. As the spectrum of the scalar sector is not very well known, and a
strong mixing between gluonic and mesonic states is present, the analysis is
however expected to be more involved and will be left for the future. Still,
as for example the recent dispersive analysis \cite{mou11} of the scalar $I=0$
sector demonstrates, the situation is not completely hopeless. Additionally,
the LET may open the route to a lattice determination of the gluon condensate.

\section{Conclusions}

The investigation of the scalar gluonium correlation function has several
interesting theoretical aspects. The relevant interpolating current represents
the gluon term in the QCD Lagrangian and its renormalisation group invariant
counterpart is proportional to the gluonic piece in the QCD energy-momentum
tensor. Furthermore, a low-energy theorem for the correlator provides a
relation to the gluon condensate, an important parameter in the framework of
QCD sum rules.

The general structure of the perturbative expansion for the scalar gluonium
correlator was discussed in section~2, and explicit analytical results are
available up to next-next-to-leading order. Furthermore, the physical
correlation function $D_{G^2}(s)$ was defined, which is the analog of the QCD
Adler function in the case of the vector current correlator. It turns out that
the perturbative corrections for $D_{G^2}(s)$ are rather large, for which
reason a phenomenological analysis of this correlator would appear questionable,
leaving aside the additional complication of the sparse knowledge on the
physical spectrum.

Still, from the theoretical perspective a general analysis of the behaviour of
the perturbative expansion, also at large orders is interesting, because on
the one hand it sheds light on the asymptotic nature of the perturbative
series, and on the other hand the singularity structure of the Borel transform
of $D_{G^2}(s)$ in the positive Borel plane is related to higher-dimensional
operator corrections in the operator product expansion.

The all-order perturbative behaviour of current correlators can be studied
in the large-$n_f$, or relatedly, the large-$\beta_0$ approximation. To this
end in section~3 the Borel transform $B[D_{G^2}](u)$ was calculated, from
which it is a simple matter to derive the large-$\beta_0$ approximation of
the correlator $D_{G^2}(s)$. Like in the case of the QCD Adler function, the
large-order behaviour is governed by the leading ultraviolet renormalon
singularity at $u=-1$, and in the $\MSb$ scheme dominance of this contribution
takes over at about the 12th perturbative order.

Contrary to the Adler function, $D_{G^2}(s)$ does not possess an infrared
renormalon pole at $u=2$, because the gluon condensate only appears in the
subtraction constant $\wh\Pi_{G^2}(0)$, which due to the derivatives vanishes
for the physical correlator $D_{G^2}(s)$. This statement should also hold true
in full QCD. Furthermore, in the large-$\beta_0$ approximation the leading
infrared renormalon singularity at $u=3$ is only logarithmic, and it is
observed that at no order this renormalon pole provides the dominant
contribution to a certain perturbative coefficient. The divergent structure is,
however, different in full QCD, for which the power-like renormalon singularity
at $u=3$ was explicitly derived from renormalisation group considerations.

An interesting limiting case are the low energy theorems of eqs.~\eqn{PiG2LET}
and \eqn{PithetaLET}, since they are solely expressed in terms of the
renormalisation group invariant gluon condensate. This does, however, not imply
that the vacuum expectation value of the (even RGI) gluonic operator is a
physical quantity. Like in the case of the pseudoscalar correlation function,
the subtraction constant is not directly related to the physical spectrum and
is expected to depend on the short-distance renormalisation. Still, the
low-energy theorems may open additional routes to extract a phenomenological
value for the gluon condensate.

These considerations bear some relevance for the question of the QCD
contribution to the vacuum energy, as such a contribution is given by the
vacuum expectation value of the trace of the energy-momentum tensor. From
the discussion above it appears as if at least in pure QCD this contribution
is unphysical, as it depends on the renormalisation prescription. An
immediate interesting question arises: what happens if QCD is coupled to
gravity? As the expectation value of the energy-momentum tensor between
particle states is physical, and a gravitational field can change particle
number, then also its vacuum expectation value should become physical. The
investigation of this question is beyond the scope of the present article,
but should be pursued in the future.

\bigskip
\acknowledgments
The author should like to thank Renata~Jora for collaboration at an early stage
of this work. Interesting discussions with Martin~Beneke, Rafel~Escribano and
Antonio~Pineda are also gratefully acknowledged. This work has been
supported in parts by the Spanish Consolider-Ingenio 2010 Programme CPAN
(CSD2007-00042) and by CICYT-FEDER FPA2008-01430 as well as FPA2011-25948.

\begin{boldmath}
\section*{Appendix A: Large-$\beta_0$ approximation for $\Pi_{G^2}(s)$}
\end{boldmath}

\addcontentsline{toc}{section}
{Appendix A: Large-\boldmath{$\beta_0$} approximation for
 \boldmath{$\Pi_{G^2}(s)$}}
\newcounter{alpha1} \renewcommand{\thesection}{\Alph{alpha1}}
\renewcommand{\theequation}{\Alph{alpha1}.\arabic{equation}}
\renewcommand{\thetable}{\Alph{alpha1}.\arabic{table}}
\setcounter{alpha1}{1} \setcounter{equation}{0} \setcounter{table}{0}

In the following, the large-$\beta_0$ approximation for the correlator
$\Pi_{G^2}(s)$ of eq.~\eqn{PiG2} will be provided and discussed. The
expression which arises from the explicit computation of the double
bubble-chain diagrams in a general space-time dimension $D=4-2\ve$ reads
\begin{eqnarray}
\label{PiG2lb0}
\Pi_{G^2}^{{\rm large-}\beta_0}(s) &=& \frac{(N_c^2-1)}{2\pi^2}\,s^2\,
(3-2\ve)\left(\frac{4\pi\mu^2}{-s}\right)^{\!\ve} \sum\limits_{m=0}^\infty
\,\sum\limits_{n=0}^\infty \left(-\,\frac{\beta_1 a_\mu}{2\,\hat\ve}
\right)^{\!m+n}\cdot \nn \\
\mvs
&& \cdot \sum\limits_{k=0}^m \,\sum\limits_{l=0}^n \binom{m}{k} \binom{n}{l}
\left[\, -\,6\,B(2-\ve,2-\ve)\,\Gamma(1+\ve)\left(\frac{4\pi\mu^2}{-s}
\right)^{\!\ve}\,\right]^{k+l}\cdot \nn \\
\mvs
&& \cdot \Big[\, 2 - (3+k+l)\ve + (k+1)(l+1)\ve^2 \,\Big]\, G_{kl}(\ve) \,,
\end{eqnarray}
where $1/\hat\ve\equiv 1/\ve-\gamma_E+\ln(4\pi)$, $B(x,y)$ is Euler's beta
function, and the function $G_{kl}(\ve)$ is found to be:
\begin{equation}
G_{kl}(\ve) \,=\, \frac{\Gamma[2-(k+1)\ve]\,\Gamma[2-(l+1)\ve]}
{\Gamma[1+k\,\ve]\,\Gamma[1+l\,\ve]\,\Gamma[4-(k+l+2)\ve]}\,
\Gamma[-2+(k+l+1)\ve] \,.
\end{equation}
The independent coefficients $c_{n,1}^{l\beta_0}$ in the large-$\beta_0$
approximation can then be obtained by working out the terms linear in the
logarithm $L=\ln(-s/\mu^2)$ from eq.~\eqn{PiG2lb0}.

In order to be able to compare to the large-$\beta_0$ approximation for the
function $D_{G^2}(s)$ of eq.~\eqn{Dlargeb0}, one also requires the coefficients
$c_{n,2}^{l\beta_0}$ and $c_{n,3}^{l\beta_0}$, that is the terms quadratic and
cubic in $L$. These can either be calculated from a direct expansion of
\eqn{PiG2lb0}, or through the following relations, valid in the large-$\beta_0$
approximation, which are derivable from the RGE:
\begin{eqnarray}
c_{n,2}^{l\beta_0} &=& -\,\frac{(n+1)}{4}\,\beta_1\,c_{n-1,1}^{l\beta_0} \,, \\
\mvs
c_{n,3}^{l\beta_0} &=& \frac{1}{12}\,\Big[\frac{n+1}{2}\Big]\left(2 \left[
\frac{n}{2}\right]+1\right) \beta_1^2 \,c_{n-2,1}^{l\beta_0} \,,
\end{eqnarray}
where $[r]$ denotes the integer part of $r$. Employing all results, one can
verify that the relation
\begin{equation}
\frac{(n+1)!}{2^n}\,\beta_1^n\,a_n \,=\, c_{n,1}^{l\beta_0} +
3\,c_{n,2}^{l\beta_0} + 3\,c_{n,3}^{l\beta_0}
\end{equation}
is satisfied, which provides a good check of the computation.

A final remark concerns the Borel transform of $\Pi_{G^2}(s)$. Through a
factor $\Gamma(u-2+\ve)$, at $D=4$ it contains UV renormalon singularities for
all integer $u\leq 2$. The ambiguity inflicted by the rightmost singularity at
$u=2$ corresponds to the corresponding one of the gluon condensate in the
subtraction constant $\Pi_{G^2}(0)$. The UV renormalon for the gluon condensate at $u=2$ has also been obtained in ref.~\cite{gs05}. Considering in addition
the momentum dependence, and using the relation
\begin{equation}
-\,s\,\frac{{\rm d}^3}{{\rm d}s^3}\,s^{2-u-\ve}\,\Gamma(u-2+\ve) \,=\,
s^{-u-\ve}\,\Gamma(1+u+\ve) \,,
\end{equation}
it becomes clear that at $D=4$ the Borel transform of the physical correlator
$D_{G^2}(s)$ only has UV renormalon singularities for $u\leq -1$, as discussed
in section~\ref{sect3}.

\bigskip
\providecommand{\href}[2]{#2}\begingroup\raggedright\endgroup

\end{document}